\documentclass[12pt,epsf,amssymb]{article} \textheight =23 truecm
\def\lsim{\raise0.3ex\hbox{$<$\kern-0.75em\raise-1.1ex\hbox{$\sim$}}}
\def\gsim{\raise0.3ex\hbox{$>$\kern-0.75em\raise-1.1ex\hbox{$\sim$}}}
\def\noi{\noindent} \def\nn{\nonumber} \def\bea{\begin{eqnarray}}
\def\eea{\end{eqnarray}} \def\beq{\begin{equation}}
\def\eeq{\end{equation}} 
\def\beeq{\begin{eqnarray}} \def\eeeq{\end{eqnarray}} \def\R{ {\rm R
\kern -.31cm I \kern .15cm}} \def\C{ {\rm C \kern -.15cm \vrule
width.5pt \kern .12cm}} \def\Z{ {\rm Z \kern -.27cm \angle \kern
.02cm}} \def\N{ {\rm N \kern -.26cm \vrule width.4pt \kern .10cm}}
\def\1{{\rm 1\mskip-4.5mu l} }

\usepackage{graphics} \usepackage{graphicx} \usepackage{epsfig}

\begin{document} \begin{center} 
\vbox to 1 truecm {}
{\large \bf Explicit form of the Isgur-Wise function in the BPS limit} \\

\vskip 1.5 truecm {\bf F. Jugeau}$^1$, {\bf A. Le Yaouanc}$^2$, {\bf L. Oliver}$^2$ {\bf and J.-C.
Raynal}$^2$\par \vskip 1 truecm

$^1$ {\it Instituto de F\'isica Corpuscular (IFIC)}\\
{\it 46071 Valencia, Spain}
\par \vskip 5 truemm

$^2$ {\it Laboratoire de Physique Th\'eorique}\footnote{Unit\'e Mixte de
Recherche UMR 8627 - CNRS }\\    {\it Universit\'e de Paris XI,
B\^atiment 210, 91405 Orsay Cedex, France} \end{center}

\vskip 1 truecm

\begin{abstract} 
Using previously formulated sum rules in the heavy quark limit of QCD, we demonstrate that if the slope $\rho^2 = - \xi '(1)$ of the Isgur-Wise function $\xi (w)$ attains its lower bound $\displaystyle{{3\over 4}}$, then all the derivatives $(-1)^L\xi^{(L)}(1)$ attain their lower bounds ${\displaystyle {(2L+1)!! \over 2^{2L}}}$, obtained by Le Yaouanc et al. This implies that the IW function is completely determined, given by the function $\xi (w) = \left ({\displaystyle {2 \over w+1}}\right)^{3/2}$. Since the so-called BPS condition proposed by Uraltsev implies $\rho^2 = \displaystyle{{3 \over 4}}$, it implies also that the IW function is given by the preceding expression.
\end{abstract}

\vskip 2 truecm

\noi LPT Orsay 06-20 \par 
\noi March 2006
\par \vskip 1 truecm

\noindent e-mails : jugeau@ific.uv.es,
leyaouan@th.u-psud.fr, oliver@th.u-psud.fr 
\newpage \pagestyle{plain}
\section{Introduction.} \hspace*{\parindent} 
In the leading order of the heavy quark expansion of QCD, Bjorken sum rule (SR) \cite{1r,2r} relates the slope of the elastic Isgur-Wise (IW) function $\xi (w)$, to the IW functions of the transition between the ground state $j^P = {1 \over 2}^-$ and the $j^P = {1 \over 2}^+, {3 \over 2}^+$ excited states, $\tau_{1/2}^{(n)}(w)$, $\tau_{3/2}^{(n)}(w)$, at zero recoil $w=1$ ($n$ is a radial quantum number). This SR leads to the lower bound $-\xi '(1) = \rho^2 \geq {1 \over 4}$. A new SR was formulated by Uraltsev in the heavy quark limit \cite{3r}, involving also $\tau_{1/2}^{(n)}(w)$, $\tau_{3/2}^{(n)}(w)$, that implies, combined with Bjorken SR, the much stronger lower bound
\beq
\label{1e}
\rho^2 \geq {3 \over 4}
\eeq

A basic ingredient in deriving this bound was the consideration of the non-forward amplitude $B(v_i) \to D^{(n)}(v') \to B(v_f)$, allowing for general $v_i$, $v_f$, $v'$ and where $B$ is a ground state meson. In refs. \cite{4r,5r,6r} we have developed, in the heavy quark limit of QCD, a manifestly covariant formalism within the Operator Product Expansion (OPE), using the matrix representation \cite{7r} for the whole tower of heavy meson states \cite{8r}. We did recover Uraltsev SR plus a general class of SR that allow to bound also higher derivatives of the IW function. In particular, we found two bounds for the curvature $\xi ''(1) = \sigma^2$ in terms of $\rho^2$, that imply
\beq
\label{2e}
\sigma^2 = \xi '' (1) \geq {15 \over 16}
\eeq

\noi Moreover we found also lower bounds for all higher derivatives, namely \cite{5r}
\beq
\label{3e}
(-1)^L\xi^{(L)}(1) \geq {(2L+1)!! \over 2^{2L}}
\eeq

\noi that reduce to (\ref{1e}) and (\ref{2e}) for the slope and the curvature.

The general SR obtained from the OPE can be written in the compact way \cite{4r}
\beq
\label{4e}
L_{Hadrons} (w_i, w_f, w_{if}) = R_{OPE} (w_i, w_f, w_{if})
\eeq

\noi where the l.h.s. is the sum over the intermediate $D$ states, while the r.h.s. is the OPE counterpart. This expression writes, in the heavy quark limit \cite{4r}~:
\bea
\label{5e}
&&\sum_{D=P,V}\sum_n Tr \left [ \overline{B}_f (v_f) \Gamma_f D^{(n)}(v')\right ] \ Tr \left [ \overline{D}^{(n)} (v') \Gamma_i B_i(v_i)\right ] \xi^{(n)}(w_i) \xi^{(n)}(w_f)\nn \\
&&+ \hbox{ Other excited states} = - 2 \xi (w_{if}) \ Tr \left [ \overline{B}_f (v_f) \Gamma_f P'_+ \Gamma_i B_i(v_i)\right ]
\eea

\noi where
\beq
\label{6e}
w_i = v_i \cdot v' \qquad w_f = v_f \cdot v' \qquad w_{if} = v_i \cdot v_f
\eeq

\noi $P'_+ = \displaystyle{{1 + {/\hskip - 2 truemm v}' \over 2}}$ is the positive energy projector on the intermediate $c$ quark, we assume that the IW functions are real and the $B$ meson is the pseudoscalar ground state $(j^P, J^P) = \left ( {1 \over 2}^-, 0^-\right )$, where $j$ is the angular momentum of the light cloud and $J$ the spin of the bound state. The heavy quark currents considered in the preceding expression are 
\beq
\label{7e}
\overline{h}_{v'}\Gamma_i h_{v_i} \qquad\qquad  \overline{h}_{v_f}\Gamma_f h_{v'}
\eeq

\noi $B(v)$, $D(v)$ are the $4 \times 4$ matrices representing the $B$, $D$ states \cite{7r,8r}, and $\overline{B} = \gamma^0B^+\gamma^0$ denotes the Dirac conjugate matrix. \par

The domain for the variables $(w_i, w_f, w_{if})$ is \cite{4r}~:
$$w_i \geq 1 \qquad\qquad w_f \geq 1$$
\beq
\label{8e}
w_iw_f - \sqrt{(w_i^2-1)(w_f^2-1)} \leq w_{if} \leq w_iw_f + \sqrt{(w_i^2 - 1) (w_f^2 -1)}
\eeq

\noi For $w_i = w_f = w$, the domain becomes
\beq
\label{9e}
w \geq 1 \qquad\qquad 1 \leq w_{if} \leq 2w^2-1
\eeq

In \cite{4r} the following SR were established. Taking $\Gamma_i = {/\hskip-2 truemm v}_i$ and $\Gamma_f = {/\hskip-2 truemm v}_f$ and $w_i = w_f = w$ one finds the so-called Vector SR
$$(w+1)^2 \sum_{L\geq 0} {L+1 \over 2L+1}\ S_L (w,w_{if}) \sum_n \left [ \tau_{L+1/2}^{(L)(n)}(w)\right ]^2+ \sum_{L\geq 1}  S_L (w,w_{if}) \sum_n \left [ \tau_{L-1/2}^{(L)(n)}(w)\right ]^2$$
\beq
\label{10e}
= \left ( 1 + 2w + w_{if}\right ) \xi (w_{if})
\eeq

\noi and for $\Gamma_i = {/\hskip-2 truemm v}_i\gamma_5$ and $\Gamma_f = {/\hskip-2 truemm v}_f\gamma_5$ one finds the Axial SR

 $$\sum_{L\geq 0}  S_{L+1} (w,w_{if}) \sum_n \left [ \tau_{L+1/2}^{(L)(n)}(w)\right ]^2 +(w-1)^2 \sum_{L\geq 1}  {L \over 2L-1} S_{L-1} (w,w_{if}) \sum_n \left [ \tau_{L-1/2}^{(L)(n)}(w)\right ]^2$$
\beq
\label{11e} 
=-  \left ( 1 - 2w + w_{if}\right ) \xi (w_{if})
\eeq

\noi In the precedent equations the IW functions $\tau_{L\pm 1/2}^{(L)(n)}(w)$ correspond to the transitions ${1 \over 2}^- \to j = L \pm {1 \over 2}$ and the function $S_{L} (w,w_{if})$ is given by the Legendre polynomial
\beq
\label{12e}
S_L(w, w_{if}) = \sum_{0 \leq k \leq L/2} C_{L,k}\left ( w^2-1\right )^{2k} \left (w^2 - w_{if}\right )^{L-2k}
\eeq

\noi with 
\beq
\label{13e}
C_{L,k} = (-1)^k {(L!)^2 \over (2L)!}\ {(2L-2k)! \over k!(L-k)!(L-2k)!}
\eeq
\vskip 5 truemm

Differentiating $n$ times both SR (\ref{10e}), (\ref{11e}) with respect to $w_{if}$ and going to the border of the domain (\ref{9e}) $w_{if} = w = 1$, one gets, among other relations,
\beq
\label{14e}
(-1)^L \xi^{(L)} (1) = {(2L+1)!! \over 2^{2L}} + {2L+1 \over 4} (L-1) ! \sum_n \left [ \tau_{L-1/2}^{(L)(n)}(1)\right ]^2
\eeq

\noi and, due to the positivity of the second term, the bounds (\ref{3e}) follow.\par

On the other hand, Uraltsev \cite{9r} has proposed a special limit of HQET, namely the so-called BPS limit, that implies
\beq
\label{15e}
\rho^2 = {3 \over 4}
\eeq

\noi among other interesting consequences for subleading quantities. We will give below a simple derivation of this value of the slope in the BPS limit.\par

In the present paper we will demonstrate, using the above SR, that if the slope reaches its lower bound (\ref{1e}), as happens in the BPS limit, then all derivatives reach their lower bounds (\ref{3e}), i.e.
\beq
\label{16e}
\rho^2 = {3 \over 4} \quad \to \quad (-1)^L\ \xi^{(L)}(1) = {(2L+1)!! \over 2^{2L}}
\eeq

\noi Then, the Isgur-Wise function is completely fixed, namely
\beq
\label{17e}
\xi (w) = \left ( {2 \over w+1}\right )^{3/2}
\eeq

The aim of this paper is a pure mathematical one, namely the demonstration of expression (\ref{17e}) for the Isgur-Wise function $\xi (w)$ in the limit $\rho^2 = {3 \over 4}$. Although in the paper we make some phenomenological remarks, the detailed comparison of this function with data, including radiative and $1/m_Q$ corrections, has to be done in a forthcoming work.\par

We have organized the paper as follows. In Section 2 we give a brief introduction to the BPS limit. In Section 3 we give a simple demonstration that the BPS limit implies $\rho^2 = {3 \over 4}$. In Section 4 we demonstrate, using the results of \cite{5r} \cite{6r}, that $\rho^2 = {3 \over 4}$ implies that the curvature attains its lower limit (\ref{2e}), i.e. $\xi ''(1) = \sigma^2 = {15 \over 16}$. Then, in Section 5 we demonstrate by induction, using the sum rules (\ref{10e}) and (\ref{11e}) that all derivatives attain their lower limit (\ref{3e}), $(-1)^L \xi^{(L)}(1) = {(2L+1)!! \over 2^{2L}}$ and therefore that the IW function has the form (\ref{17e}). In Section 6 we conclude and briefly recall other consequences of the BPS limit for subleading form factors. In Appendix A we give an alternative derivation of the value of the curvature in the BPS limit, making explicit use of the BPS hypothesis to illustrate some interesting physical features.

\section{The BPS limit.}
 \hspace*{\parindent} The motivation to introduce the BPS limit \cite{9r} has been the rather close values obtained from experiment in inclusive $B$ decay for the fundamental parameters $\mu_{\pi}^2$ and $\mu_G^2$~: 
\beq
\label{18e}
\mu_{\pi}^2 =\ - {<B(v)|O_{kin, v}^{(b)}|B(v)>\over 2m_B} \qquad\quad \mu_G^2 =\ {<B(v)|O_{mag, v}^{(b)}|B(v)> \over 2m_B}
\eeq

\noi i.e. the matrix elements of the operators that appear in the $1/m_Q$ perturbation of the HQET Lagrangian,
\bea
\label{19e}
&{\cal L}_{kin,v}^{(Q)} = \displaystyle{{1 \over 2m_Q}}\ O_{kin,v}^{(Q)} &\qquad\qquad {\cal L}_{mag, v}^{(Q)} = {1 \over 2m_Q}\ O_{mag, v}^{(Q)}\\
&\quad O_{kin,v}^{(Q)} = \overline{h}_v^{(Q)}(iD)^2 h_v^{(Q)} &\qquad\qquad O_{mag,v}^{(Q)} = {g_s \over 2} \ \overline{h}_v^{(Q)}  \sigma_{\alpha\beta} G^{\alpha\beta}h_v^{(Q)}
\label{20e}
\eea

\noi In terms of ${1 \over 2}^- \to {1 \over 2}^+$, ${3 \over 2}^+$ Isgur-Wise functions at zero recoil $\tau_j^{(n)}(1)$ and level spacings $\Delta E_j^{(n)}$ $(j = {1 \over 2}, {3 \over 2})$, these quantities read \cite{10r}
\bea
\label{21e}
&&\mu_{\pi}^2 = 6 \sum_n \left [ \Delta E_{3/2}^{(n)}\right ]^2 \left [ \tau_{3/2}^{(n)}(1)\right ]^2 + 3 \sum_n \left [ \Delta E_{1/2}^{(n)}\right ]^2 \left [ \tau_{1/2}^{(n)}(1)\right ]^2\\
&&\mu_{G}^2 = 6 \sum_n \left [ \Delta E_{3/2}^{(n)}\right ]^2 \left [ \tau_{3/2}^{(n)}(1)\right ]^2 - 6 \sum_n \left [ \Delta E_{1/2}^{(n)}\right ]^2 \left [ \tau_{1/2}^{(n)}(1)\right ]^2
\label{22e}
\eea

\noi The inequality $\mu_{\pi}^2 \geq \mu_G^2$ holds, and one has found empirically, from the inclusive decay  $\overline{B}_d \to X_c \ell \overline{\nu}_{\ell}$, that $\mu_{\pi}^2$ and $\mu_G^2$ are rather close \cite{11r}
\beq
\label{23e}
\mu_{\pi}^2 \cong 0.4 \ {\rm GeV}^2\qquad\qquad\qquad   \mu_G^2 \cong 0.35\ {\rm GeV}^2
\eeq

Uraltsev has suggested a dynamical hypothesis that implements the limiting condition of $\mu_{\pi}^2$ and $\mu_G^2$ being equal, the so-called BPS approximation,
\beq
\label{24e}
\mu_{\pi}^2 = \mu_G^2
\eeq

The values quoted in (\ref{23e}) deserve an important comment. Both matrix elements are not on equal footing. The value of $\mu_G^2 \cong 0.35$~GeV$^2$ is obtained from the heavy-light mesons hyperfine splitting (see for example ref. \cite{9r}), while the value $\mu_{\pi}^2 \cong 0.4$~GeV$^2$ comes from the fit to inclusive $\overline{B}_d \to X_c \ell \overline{\nu}_\ell$ decay moments. These are the central values of the starting expression (2) of ref. \cite{11r} leading to the inclusive determination of $|V_{cb}|$. However the final fitted values in inclusive decays \cite{11r} turn out to be $\mu_G^2 \cong 0.297$~GeV$^2$ and $\mu_{\pi}^2 \cong 0.401$~GeV$^2$ (Table 2 of ref. \cite{11r}). On the other hand, strictly speaking, these parameters are scale dependent~: $\mu_G^2(\mu )$ and $\mu_{\pi}^2 (\mu )$. We do not discuss here the radiative corrections, and this $\mu$-dependence is not relevant within the theoretical framework adopted in this paper. The values given above correspond to $\mu \cong 1$~GeV \cite{11r}.\par

The discrepancy between the hyperfine and the semileptonic determinations of $\mu_G^2$ is a problem that should have its answer either in further experimental data or on a careful HQET analysis of both determinations, that are quite different on physical grounds. The hyperfine splitting determination of $\mu_G^2$ seems to us a more direct and reliable one, and we have adopted the values (\ref{23e}), that were at the basis of the BPS limit proposed by Uraltsev \cite{9r}. However, it could turn out that the BPS approximation is not as good as it seems, as one sees adopting the values $\mu_G^2 \cong 0.297$~GeV$^2$ and $\mu_{\pi}^2 \cong 0.401$~GeV$^2$.\par

Let us underline that the main purpose of this paper is a mathematical one within the heavy quark limit of QCD. Namely, the determination of the form of the Isgur-Wise function in the heavy quark limit by adding one dynamical assumption, the BPS condition.\par

 Let us consider the pseudoscalar $B$ meson at rest, $v = (1,0,0,0)$. The equation of motion of HQET in the heavy quark limit implies
\beq
\label{25e}
iD^0h_v^{(b)}|B(v)>\ = 0
\eeq

\noi where $D^{\mu}$ is the covariant derivative and $h_v$ is the heavy quark field.\par

Uraltsev has proposed a new more specific constraint, valid only for the {\it pseudoscalar ground state} meson $B$, the so-called BPS constraint
\beq
\label{26e}
\left ( \vec{\sigma}\cdot i \overrightarrow{D}\right ) h_v^{(b)}|B(v)>\ = 0
\eeq

\noi that amounts to the vanishing of the smaller components of the heavy quark field {\it within the pseudoscalar} $B$ meson.\par

It will be convenient in the following to write these two conditions in a covariant way, for any value of $v$. These equations then read,
\beq
\label{27e}
(iD\cdot v)h_v^{(b)}|B(v)>\ = 0
\eeq
\beq
\label{28e}
\gamma_5 i {/\hskip - 3 truemm D} h_v^{(b)}|B(v) >\ = 0
\eeq

\noi From the identity
\beq
\label{29e}
i  {/\hskip - 3 truemm D}  i  {/\hskip - 3 truemm D}  = (iD)^2 + {g_s \over 2}  \sigma_{\alpha\beta} G^{\alpha\beta}
\eeq

\noi one observes that (\ref{28e}) implies the equality (\ref{24e}).

\section{The BPS limit implies ${\bf \rho^2 = {3 \over 4}}$.}
\hspace*{\parindent} Let us now choose the transition $\left ( {1 \over 2}^-, 0^-\right ) \to \left ( {1 \over 2}^+, 0^+\right )$ (where any radial excitation $n$ can be assumed) and consider the matrix elements defined by Leibovich et al. \cite{12r},
\beq
\label{30e}
<D(0^+)(v') | \overline{h}_{v'}^{(c)}\Gamma h_v^{(b)}|B(v)>\ = 2 \tau_{1/2}(w) Tr \left [P'_+\Gamma P_+(-\gamma_5)\right ]
\eeq

\noi and
\bea
\label{31e}
&&<D(0^+)(v') | \overline{h}_{v'}^{(c)}\Gamma i \overrightarrow{\cal D}_{\lambda} h_v^{(b)}|B(v)>\ = Tr \left [S_{\lambda}^{(b)} P'_+\Gamma P_+(-\gamma_5)\right ]\nn \\
&&<D(0^+)(v') | \overline{h}_{v'}^{(c)} i \overleftarrow{\cal D}_{\lambda} \Gamma h_v^{(b)}|B(v)>\ = Tr \left [ S_{\lambda}^{(c)} P'_+\Gamma P_+(-\gamma_5)\right ]
\eea

\noi where $P_+ = \displaystyle{{1 + {/\hskip-2 truemm v} \over 2}}$, $n$ is a radial quantum number, $\Gamma$ is any Dirac matrix and $S_{\lambda}^{(Q)}$ reads 
\beq
\label{32e}
S_{\lambda}^{(Q)} = \zeta_1^{(Q)} v_{\lambda} + \zeta_2^{(Q)} v'_{\lambda} + \zeta_3^{(Q)} \gamma_{\lambda}
\eeq

\noi The equations of motion for the heavy quark $b$ and $c$ imply respectively
\bea
\label{33e}
&&Tr \left [ S_{\lambda}^{(b)} P'_+\Gamma \gamma_5 v^{\lambda} P_+(-\gamma_5)\right ] \nn \\
&&= Tr \left \{ \left [  \zeta_1^{(b)} v_{\lambda} + \zeta_2^{(b)} v'_{\lambda} + \zeta_3^{(b)} \gamma_{\lambda}\right ] P'_+\Gamma \gamma_5 v^{\lambda} P_+(-\gamma_5)\right \} = 0
\eea
\bea
\label{34e}
&&Tr \left [ S_{\lambda}^{(c)} P'_+\Gamma \gamma_5 v'^{\lambda} P_+(-\gamma_5)\right ] \nn \\
&&= Tr \left \{ \left [  \zeta_1^{(c)} v_{\lambda} + \zeta_2^{(c)} v'_{\lambda} + \zeta_3^{(c)} \gamma_{\lambda}\right ] P'_+\Gamma \gamma_5 v'^{\lambda} P_+(-\gamma_5)\right \} = 0
\eea

\noi i.e.
\bea
&&\zeta_1^{(b)}(w) + w\zeta_2^{(b)}(w) - \zeta_3^{(b)}(w) = 0 \nn \\
&&w\zeta_1^{(c)}(w) + \zeta_2^{(b)}(w) + \zeta_3^{(b)}(w) = 0 
\label{35e}
\eea

\noi On the other hand, from translational invariance,
\bea
\label{36e}
&&i \partial_{\lambda} <D(0^+) (v')| \overline{h}_{v'}^{(c)}\Gamma h_v^{(b)}|B(v)>\nn\\
&&=\ <D(0^+)(v') | \overline{h}_{v'}^{(c)}(\Gamma i \overrightarrow{\cal D}_{\lambda} + i \overleftarrow{\cal D}_{\lambda} \Gamma) h_v^{(b)}|B(v)>\nn \\
&&= \left ( \overline{\Lambda} v_{\lambda} - \overline{\Lambda}^* v'_{\lambda}  \right ) Tr \left [ P'_+\Gamma P_+(-\gamma_5)\right ]2 \tau_{1/2}(w) 
\eea

\noi and therefore
\beq
\label{37e}
S_{\lambda}^{(b)} + S_{\lambda}^{(c)} = \left ( \overline{\Lambda} v_{\lambda} - \overline{\Lambda}^* v'_{\lambda}  \right ) 2 \tau_{1/2}(w) 
\eeq

\noi where $\overline{\Lambda}$ and $\overline{\Lambda}^*$ are the energies of the light degrees of freedom.\par

As demonstrated in \cite{12r}, the equations of motion that imply (\ref{35e}), together with (\ref{36e}), imply, writing for any $n$,
\beq
\label{38e}
\zeta_3^{(b)(n)}(1) = - {1 \over 2} \Delta E_{1/2}^{(n)} \tau_{1/2}^{(n)}(1)
\eeq

\noi Let us now apply the BPS condition (\ref{28e}), that acts only on the $b$ quark~:
\beq
\label{39e}
<D(0^+)(v') | \overline{h}_{v'}^{(c)}\Gamma \gamma_5i {/\hskip-3 truemm D} h_v^{(b)}|B(v)> = 0
\eeq

\noi that gives
\beq
\label{40e}
Tr \left [ S_{\lambda}^{(b)} P'_+\Gamma \gamma_5 \gamma^{\lambda} P_+(-\gamma_5)\right ] = Tr \left \{ \left [  \zeta_1^{(b)} v_{\lambda} + \zeta_2^{(b)} v'_{\lambda} + \zeta_3^{(b)} \gamma_{\lambda}\right ] P'_+\Gamma \gamma_5 \gamma^{\lambda} P_+(-\gamma_5)\right \} = 0
\eeq

\noi or
\beq
\label{41e}
\left [  \zeta_1^{(b)}  + \zeta_2^{(b)} + \zeta_3^{(b)}\right ] Tr \left [ P'_+ \Gamma {/ \hskip - 2 truemm v}\right ] - \left [  \zeta_1^{(b)}  + (2w-1)  \zeta_2^{(b)} - 2 \zeta_3^{(b)}\right ]  Tr \left [ P'_+ \Gamma \right ] = 0
\eeq

\noi Since the matrix $\Gamma$ is arbitrary, one has
\bea
\label{42e}
&&\zeta_1^{(b)}(w)  +  \zeta_2^{(b)}(w) + \zeta_3^{(b)}(w) = 0 \nn \\
&&\zeta_1^{(b)}(w)  +  (2w-1) \zeta_2^{(b)}(w) -2 \zeta_3^{(b)}(w) = 0
\eea

\noi that implies, at zero recoil
\beq
\label{43e}
\zeta_1^{(b)}(1)  +  \zeta_2^{(b)}(1) = \zeta_3^{(b)}(1) = 0
\eeq

\noi Combining with (\ref{38e}) one has, for all $n$~:
\beq
\label{44e}
\tau_{1/2}^{(1)(n)}(1) = 2 \tau_{1/2}^{(n)}(1) = 0
\eeq

\noi in terms of the notation $\tau_{1/2}^{(1)(n)}(w) = 2 \tau_{1/2}^{(n)}(w)$.\par

To summarize, this relation follows from translational invariance and the equations of motion plus the BPS condition.\par

On the other hand, (\ref{14e}) reads for $L = 1$
\beq
\label{45e}
\rho^2 = {3 \over 4} + {3 \over 4} \sum_n \left [ \tau_{1/2}^{(1)(n)}(1)\right ]^2
\eeq

\noi Therefore, relation (\ref{44e}) implies (\ref{15e}), $\rho^2 = {3 \over 4}$.\par

The result (\ref{44e}) is very strong. Uraltsev sum rule \cite{3r} reads,
\beq
\label{NEW46}
{1 \over 3} \sum_n \left [ \tau_{3/2}^{(1)(n)}(1) \right ]^2 - {1 \over 4} \sum_n \left [ \tau_{1/2}^{(1)(n)}(1) \right ]^2 = {1 \over 4}
\eeq

\noi and is therefore consistent with the BPS approximation, that implies (\ref{44e}), and therefore $\sum\limits_n [\tau_{3/2}^{(1)}(1)]^2 = {3 \over 4}$.\par

However, on experimental grounds, and keeping only the $n=0$ states, for which there are some experimental indications, the relation (\ref{44e}) does not seem satisfied.\par

The experimental situation is very involved, as pointed out in detail in \cite{13rnew} and \cite{14rnew}. In these papers the different experimental data are discussed, some of which even violate Uraltsev SR, and the predictions for $\tau_{1/2}^{(1)(0)}(w)$, $\tau_{1/2}^{(1)(0)}(w)$ in relativistic quark models, QCD Sum Rules and Lattice QCD are given. These do not point out to the relation (\ref{44e}) for $n=0$, but some schemes satisfy Uraltsev SR, giving (\ref{NEW46}) approximately saturated by the $n=0$ states.\par

On the other hand, if (\ref{24e}) turns out eventually to be approximately satisfied on empirical grounds in both hyperfine and semileptonic inclusive data, the question remains of the comparison of the $\overline{B}_d\to D^{**}\ell \overline{\nu}_{\ell}$ Isgur-Wise functions and level spacings and the values (\ref{23e}) for $\mu_G^2$ and $\mu_{\pi}^2$ with the SR (\ref{21e}) and (\ref{22e}).\par

To summarize, in data as well as in the different dynamical schemes, relation (\ref{44e}) does not hold. However, theoretically one expects (\ref{NEW46}) to hold approximately for $n = 0$ states, as is satisfied in some dynamical schemes. Relation (\ref{44e}) is just one mathematical limit satisfying (\ref{24e}), and approximately satisfying (\ref{23e}).

\section{Curvature of the IW function in the limit ${\bf \rho^2 = {3 \over 4}}$.}
\hspace*{\parindent} Here we will start from (\ref{15e}) or (\ref{44e}) and demonstrate that it implies that the curvature (\ref{2e}) attains its lower bound,
\beq
\label{46e}
\sigma^2 = \xi '' (1) = {15 \over 16}
\eeq

To proof (\ref{46e}) from (\ref{44e}) we use the sum rules in the heavy quark limit (\ref{10e}) and (\ref{11e}) obtained in \cite{5r}, differentiating relatively to $w_{if}$ and to $w$ and going to the border of the domain (\ref{9e}) $w_{if} = w = 1$ (formulas (33), (34), (44), (47) and (48) of \cite{6r}),
\beq
\label{47e}
\rho^2 - 2 \sigma^2 + {12 \over 5} \sum_n \left [ \tau_{5/2}^{(2)(n)}(1)\right ]^2 + \sum_n \left [ \tau_{3/2}^{(2)(n)}(1)\right ]^2  = 0
\eeq
\bea
\label{48e}
&&\rho^2 - {4 \over 3} \sum_n \left [ \tau_{3/2}^{(1)(n)}(1)\right ]^2 - {8 \over 3} \sum_n \tau_{3/2}^{(1)(n)}(1) \tau_{3/2}^{(1)(n)'}(1) -  \sum_n \tau_{1/2}^{(1)(n)}(1) \tau_{1/2}^{(1)(n)'}(1)\nn \\
&&-2\sum_n \left [ \tau_{3/2}^{(2)(n)}(1)\right ]^2 - {24 \over 5} \sum_n  \left [ \tau_{5/2}^{(2)(n)}(1)\right ]^2 = 0
\eea
\beq
\label{49e}
\rho^2 = \sum_n \left [ \tau_{3/2}^{(1)(n)}(1)\right ]^2 
\eeq
\beq
\label{50e}
\sigma^2 = 2 \sum_n \left [ \tau_{5/2}^{(2)(n)}(1)\right ]^2 
\eeq
\beq
\label{51e}
\sigma^2 = 2 \sum_n \tau_{3/2}^{(1)(n)}(1) \tau_{3/2}^{(1)(n)'}(1) + 6 \sum_n \left [ \tau_{5/2}^{(2)(n)}(1)\right ]^2
\eeq

\noi From (\ref{44e}), equation (\ref{48e}) becomes
\bea
\label{52e}
&&\rho^2 - {4 \over 3} \sum_n \left [ \tau_{3/2}^{(1)(n)}(1)\right ]^2 - {8 \over 3} \sum_n \tau_{3/2}^{(1)(n)}(1) \tau_{3/2}^{(1)(n)'}(1)\nn \\
&&-2\sum_n \left [ \tau_{3/2}^{(2)(n)}(1)\right ]^2 - {24 \over 5} \sum_n  \left [ \tau_{5/2}^{(2)(n)}(1)\right ]^2 = 0
\eea

\noi Combining (\ref{47e}), (\ref{49e}), (\ref{50e}), (\ref{51e}) and (\ref{52e}) one obtains the two equations
\bea
\label{53e}
&&\sigma^2 = {15 \over 16} + {5 \over 4}  \sum_n \left [ \tau_{3/2}^{(2)(n)}(1)\right ]^2 =0\nn \\
&&\sigma^2 = {15 \over 16} + {15 \over 2}  \sum_n \left [ \tau_{3/2}^{(2)(n)}(1)\right ]^2 =0
\eea

\noi that imply
\beq
\label{54e}
\tau_{3/2}^{(2)(n)}(1) = 0 \qquad \qquad \hbox{and} \qquad \qquad \sigma^2 = {15 \over 16}
\eeq

\noi and relation (\ref{46e}) is demonstrated.

\section{Form of the IW function for ${\bf \rho^2 = {3 \over 4}}$.}
\hspace*{\parindent} We have therefore by BPS $\rho^2 = {3 \over 4}$ for the slope, that implies $\sigma^2 = {15 \over 16}$ for the curvature. We will demonstrate now that the $L$-th derivative attains its lower bound (\ref{3e}) $(-1)^L \xi^{(L)}(1) = {(2L+1)!! \over 2^{2L}}$ in all generality. We make the proof by induction. We will assume relation (\ref{16e}) for $L-1$,
\beq
\label{55e}
(-1)^{L-1} \xi^{(L-1)}(1) = {(2L-1)!! \over 2^{2(L-1)}}
\eeq

\noi and use the SR (\ref{10e}) and (\ref{11e}) to demonstrate (\ref{16e}) for $L$.\par

Let us differentiate the SR (\ref{10e}), (\ref{11e}) $M$ times relatively to $w_{if}$. Using (\ref{12e})-(\ref{13e}), we need
\beq
\label{56e}
\left [ {\partial^M \over \partial w_{if}^M} S_L(w,w_{if})\right ]_{w_{if} = 1} = F_{L,M}(w)
\eeq

\noi where
\beq
\label{57e}
 F_{L,M}(w) =  R_{L,M}(w^2-1)^{L-M}
\eeq

\beq
\label{58e}
 R_{L,M} = (-1)^M \sum_{0\leq k \leq (L-M)/2} (-1)^k {(L!)^2 \over (2L)!}\ {(2L-2k)! \over k!(L-k)!(L-M -2k)!}
\eeq

\noi From the Vector SR (\ref{10e}) one obtains
\bea
\label{59e}
&&(w+1)^2 \sum_{L\geq 0} {L+1 \over 2L+1} F_{L,M}(w) \sum_n \left [ \tau_{L+1/2}^{(L)(n)}(w)\right ]^2\nn \\
&&+ \sum_{L\geq 1} F_{L,M}(w) \sum_n \left [ \tau_{L-1/2}^{(L)(n)}(w)\right ]^2 = 2(w+1) \xi^{(M)} (1) + M \xi^{(M-1)} (1) 
\eea
\vskip 5 truemm

\noi while from the Axial SR (\ref{11e}),
\beq
\label{60e}
\sum_{L\geq 0} F_{L+1,M}(w) \sum_n \left [ \tau_{L+1/2}^{(L)(n)}(w)\right ]^2
\eeq
$$+(w-1)^2  \sum_{L\geq 1} {L \over 2L-1}F_{L-1,M}(w) \sum_n \left [ \tau_{L-1/2}^{(L)(n)}(w)\right ]^2 = 2(w-1) \xi^{(M)} (1) - M\xi^{(M-1)}(1)$$

\vskip 5 truemm
\noi From the Vector SR (\ref{33e}) we obtain two useful relations.\par

1) Take $M = L$ and $w=1$,
\bea
\label{61e}
&&{4(L+1) \over 2L+1} L! \sum_n \left [ \tau_{L+1/2}^{(L)(n)}(1)\right ]^2 + L! \sum_n \left [ \tau_{L-1/2}^{(L)(n)}(1)\right ]^2\nn\\
&&= 4(-1)^L  \xi^{(L)} (1) - L(-1)^{L-1}  \xi^{(L-1)} (1)
\eea

2) Take $M = L-1$, differentiate once relatively to $w$ and take $w=1$,
\bea
\label{62e}
&&{L \over 2L-1} 4(L-1)! \sum_n \left [ \tau_{L-1+1/2}^{(L-1)(n)}(1)\right ]^2 + {L+1 \over 2L+1} 8L! \sum_n \left [ \tau_{L+1/2}^{(L)(n)}(1)\right ]^2\nn\\
&&+ {L \over 2L-1} 4(L-1)! \left  [ {\partial \over \partial w} \left \{ \sum_n \left [ \tau_{L-1+1/2}^{(L-1)(n)}(w)\right ]^2\right \} \right]_{w=1} + 2L! \sum_n  \left [ \tau_{L-1/2}^{(L)(n)}(1)\right ]^2\nn\\
&&+ (L-1)! \left  [ {\partial \over \partial w} \left \{ \sum_n \left [ \tau_{L-1-1/2}^{(L-1)(n)}(w)\right ]^2\right \} \right]_{w=1} = 2(-1)^{L-1} \xi^{(L-1)} (1)
\eea
\vskip 5 truemm

\noi Similarly, from the Axial SR (\ref{56e}) we obtain two other relations.\par

1) $M=L-1$ and $w=1$,
\beq
\label{63e}
(-1)^{L-1} \xi^{(L-1)}(1) = (L-1)! \sum_n \left [ \tau_{L-1+1/2}^{(L-1)(n)}(1)\right ]^2
\eeq

2) $M = L$ and $w=1$,
\beq
\label{64e}
(-1)^{L} \xi^{(L)}(1) = L! \sum_n \left [ \tau_{L+1/2}^{(L)(n)}(1)\right ]^2
\eeq

3) $M = L$, differentiate once relatively to $w$ and make $w=1$,
\bea
\label{65e}
&&(-1)^{L} \xi^{(L)}(1) = (L+1)! \sum_n \left [ \tau_{L+1/2}^{(L)(n)}(1)\right ]^2\nn \\
&&+ L!{1 \over 2} \left  [ {\partial \over \partial w} \left \{ \sum_n \left [ \tau_{L-1+1/2}^{(L-1)(n)}(w)\right ]^2\right \} \right]_{w=1} \eea

\noi Equations (\ref{61e})-(\ref{62e}) and (\ref{63e})-(\ref{65e}) are the generalizations to all $L$ of respectively equations (\ref{47e})-(\ref{48e}) and (\ref{49e})-(\ref{51e}).\par

To proceed with the proof by induction, we assume the vanishing of 
\beq
\label{66e}
\tau_{L-1-1/2}^{(L-1)(n)}(1)=0
\eeq

\noi that implies, from (\ref{14e}),
\beq
\label{67e}
(-1)^{L-1} \xi^{(L-1)}(1) = {(2L-1)!! \over 2^{2(L-1)}} 
\eeq

\noi Moreover, this implies that the following quantity appearing in expression (\ref{62e}) must vanish
\beq
\label{68e}
(L-1)! \left  [ {\partial \over \partial w} \left \{ \sum_n \left [ \tau_{L-1-1/2}^{(L-1)(n)}(w)\right ]^2\right \} \right]_{w=1} = 0 
\eeq

\noi and  (\ref{62e})  simplifies to
\bea
\label{69e}
&&{L \over 2L-1} 4(L-1)! \sum_n \left [ \tau_{L-1+1/2}^{(L-1)(n)}(1)\right ]^2 + {L+1 \over 2L+1} 8L! \sum_n \left [ \tau_{L+1/2}^{(L)(n)}(1)\right ]^2\nn\\
&&+ {L \over 2L-1} 4(L-1)! \left  [ {\partial \over \partial w} \left \{ \sum_n \left [ \tau_{L-1+1/2}^{(L-1)(n)}(w)\right ]^2\right \} \right]_{w=1} + 2L! \sum_n  \left [ \tau_{L-1/2}^{(L)(n)}(1)\right ]^2\nn\\
&&= 2(-1)^{L-1} \xi^{(L-1)} (1)
\eea

\noi From (\ref{64e})-(\ref{65e}), one gets
\beq
\label{70e}
(L-1)! \left  [ {\partial \over \partial w} \left \{ \sum_n \left [ \tau_{L-1+1/2}^{(L-1)(n)}(w)\right ]^2\right \} \right]_{w=1} =  - 2(-1)^{L} \xi^{(L)} (1)
 \eeq

\beq
\label{71e}
L!  \sum_n \left [ \tau_{L+1/2}^{(L)(n)}(1)\right ]^2=  (-1)^{L} \xi^{(L)} (1)
\eeq

\noi and using these relations together with (\ref{63e}) in the Vector SR (\ref{61e}) and (\ref{69e}) one obtains finally
\bea
\label{72e}
&& (-1)^{L} \xi^{(L)} (1) = {(2L+1)!! \over 2^{2L}} + {2L+1 \over 4L}  L! \sum_n \left [ \tau_{L-1/2}^{(L)(n)}(1)\right ]^2\nn\\
&& (-1)^{L} \xi^{(L)} (1) = {(2L+1)!! \over 2^{2L}} + {4L^2-1 \over 4}  L! \sum_n \left [ \tau_{L-1/2}^{(L)(n)}(1)\right ]^2
\eea

\noi that reduce to (\ref{53e}) for $L=2$ and imply
\beq
\label{73e}
\tau_{L-1/2}^{(L)(n)}(1) = 0 \qquad\quad \hbox{and} \qquad\quad (-1)^{L} \xi^{(L)} (1) = {(2L+1)!! \over 2^{2L}} 
\eeq

\noi as we wanted to demonstrate.\par

Since (\ref{73e}) are the successive derivatives of (\ref{17e}), assuming natural regularity properties, in the BPS limit the Isgur-Wise function is given by expression (\ref{17e}).

\section{Conclusion and prospects.}
\hspace*{\parindent} In conclusion, we have demonstrated in this paper that if the heavy quark limit of QCD is supplemented with a dynamical assumption, namely the BPS approximation proposed by Uraltsev, the Isgur-Wise function is completely determined, given by the expression
\beq
\label{75e}
\xi (w) = \left ( {2 \over w+1}\right )^{3/2}
\eeq
\vskip 5 truemm

This is a mathematical result that comes from the heavy quark limit of QCD plus the BPS condition introduced by Uraltsev. The comparison with data is not straightforward, since $1/m_Q$ and radiative corrections have not been taken into account. Indeed, the function that has to be extrapolated at $w=1$ to obtain $|V_{cb}|$ is the form factor $h_{A_1}(w)$, and moreover the two ratios of form factors $R_1(w)$, $R_2(w)$ are involved, that become $R_1(w) = R_2(w) = 1$ in the heavy quark limit, considered in this paper. In a recent BaBar paper, the fit to $h_{A_1}(w)$ gives a slope $\rho_{A_1}^2 = 1.14$ \cite{18r}. This is far away from the heavy quark limit result with the BPS condition $\rho^2 = 0.75$. However, to make a proper comparison, radiative corrections to the heavy quark plus BPS limit should be considered \cite{19rnew}, and the constraints on the slope from Voloshin SR, that result in an upper bound on $\rho^2$ that is close to the BPS limit, should also be taken into account \cite{20rnew}. This discussion deserves a delicate and detailed discussion that will be done elsewhere. But let us here below advance the results that stem from the BPS limit for the $1/m_Q$ corrections for $h_{A_1}(w)$. \par

Let us recall also that we did obtain elsewhere other interesting consequences of the BPS limit for the elastic subleading form factors at order $1/m_Q$, namely the Current perturbations $\xi_3(w)$ and $\overline{\Lambda} \xi (w)$ (in the notation of Falk and Neubert \cite{13r}) and the Lagrangian perturbations $\chi_1(w)$, $\chi_2(w)$ and $\chi_3(w)$ (in the notation of Luke \cite{14r}). \par

From the final results of ref. \cite{15r} for $\xi_3(w)$ and $\overline{\Lambda} \xi (w)$~: 
\bea
\label{76e}
&&\overline{\Lambda} \xi (w) = 2(w+1) \sum_n \Delta E_{3/2}^{(n)} \tau_{3/2}^{(n)}(1) \tau_{3/2}^{(n)}(w) + 2 \sum_n \Delta E_{1/2}^{(n)} \tau_{1/2}^{(n)}(1) \tau_{1/2}^{(n)}(w)\nn \\
&& \xi_3 (w) = (w+1) \sum_n \Delta E_{3/2}^{(n)} \tau_{3/2}^{(n)}(1) \tau_{3/2}^{(n)}(w) - 2 \sum_n \Delta E_{1/2}^{(n)} \tau_{1/2}^{(n)}(1) \tau_{1/2}^{(n)}(w)\eea

\noi one obtains, in the BPS limit, that implies $\tau_{1/2}^{(n)}(1) = 0$,
\beq
\label{77e}
\overline{\Lambda} \xi (w) = 2 \xi_3(w)
\eeq

\noi and therefore in this limit there is only one independent subleading form factor of the current type.\par

On the other hand, in \cite{16r} we did obtain from bounds on the $1/m_Q$ Lagrangian perturbations that in the limit in which the slope and curvature of the elastic IW function tend to their BPS values $\rho^2 \to {3 \over 4}$, $\sigma^2 \to {15 \over 16}$ one gets at zero recoil,
\beq
\label{78e}
\chi'_1(1) = \chi_2(1) = \chi'_3(1) = 0
\eeq

\noi However, for these corrections, unlike (\ref{77e}), we did not obtain results for all $w$.\par

The conditions (\ref{77e}) and (\ref{78e}) are consistent with the claim by Uraltsev that for the decay $\overline{B}_d \to D \ell \overline{\nu}_{\ell}$ there is a proportionality in the BPS limit between the two form factors $f_+(q^2)$ and $f_0 (q^2)$ \cite{9r}.\par

The results (\ref{75e}), (\ref{77e}) and (\ref{78e}) are strong constraints on the behaviour of the form factors up to order $1/m_Q$ included. An important example is the axial form factor $h_{A_1}(w)$ that enters in the $\overline{B}_d \to D^* \ell \overline{\nu}_{\ell}$ differential rate near zero recoil. From the expression for $h_{A_1}(w)$ up to order $1/m_Q$ \cite{13r} \cite{14r},
\bea
\label{79e}
&&h_{A_1}(w) = \xi (w) + {1 \over 2m_c} \left \{ \left [ 2 \chi_1(w) - 4 \chi_3(w) \right ] + {w-1 \over w+1} \ \overline{\Lambda} \xi (w)\right \}\nn \\
&&+ {1 \over 2 m_b} \left \{ \left [ 2 \chi_1(w) - 4(w-1) \chi_2(w) + 12 \chi_3(w)\right ]  - {w-1 \over w+1} \left [ - \overline{\Lambda} \xi (w) + 2 \xi_3(w) \right ] \right \}\nn \\
&&+ \ O\left  ( 1/m_Q^2 \right )
\eea

\noi one finds from (\ref{77e}) and (\ref{78e}) that the slope $h'_{A_1}$ is given in the BPS limit by the expression
\beq
\label{80e}
- h'_{A_1} (1) = {3 \over 4} - {\overline{\Lambda}  \over 4m_c}
\eeq

\noi This result gives a slope that is much smaller than the fitted values for $- h'_{A_1}(1)$  \cite {18r}. Therefore, the prospect that the BPS limit is a good approximation for exclusive semileptonic decays does not seem very good, although one cannot still draw a definite conclusion, in view of the problems outlined above and below. \par

In conclusion, we have obtained an explicit expression for the Isgur-Wise function $\xi (w)$ by implementing the heavy quark limit of QCD with a dynamical assumption, namely the BPS condition proposed by Uraltsev, coming from the condition $\mu_G^2 = \mu_{\pi}^2$ or, equivalently, from $\rho^2 = {3 \over 4}$.\par

In the comparison of these clear-cut results with experiment a number of miscellaneous problems remain open~:\par

1) There seems to be some discrepancy between the hyperfine splitting determination of $\mu_G^2 \cong 0.35$~GeV$^2$ and the best fit to the semileptonic inclusive data $\mu_G^2 \cong 0.30$~GeV$^2$. This point should be settled.\par

2) The degree of validity of the BPS approximation should be studied and compared both in the exclusive and in the inclusive decays. In particular, if one confirms the values (\ref{23e}) for $\mu_{\pi}^2$ and $\mu_G^2$, a puzzle could arise, namely why the BPS limit seems a good approximation in the inclusive decays $\overline{B}_d \to X_c \ell \overline{\nu}_{\ell}$ and less so in the exclusive case, in view of (\ref{80e}). \par

3) For the moment, all experiments do not satisfy Uraltsev SR (\ref{NEW46}), keeping the $n = 0$ states. On the other hand, some dynamical theoretical schemes do verify it, as discussed in detail in \cite{13rnew}  \cite{14rnew}. If, eventually, data do converge to a satisfaction of Uraltsev SR, one could consider as an expansion parameter around the BPS condition the dimensionless quantity arising from relation (\ref{45e}) \cite{9r}
\beq
\label{82NEW}
\rho^2 - {3 \over 4} = {3 \over 4} \sum_n \left [ \tau_{1/2}^{(1)(n)}(1) \right ]^2
\eeq

\noi The difference $\mu_{\pi}^2 - \mu_G^2$ does not seem so suited for this purpose, since this is a dimensionful quantity and poses the question of the scale on the denominator of this difference \cite{21rnew}.\par

4) The radiative corrections and $1/m_Q$ corrections should be taken into account in all form factors in the BPS limit, considering {\it both} exclusive channels $\overline{B}_d \to D \ell \overline{\nu}_{\ell}$ and $\overline{B}_d \to D^* \ell \overline{\nu}_{\ell}$.\par

5) Last but not least, the constraint from Voloshin SR \cite{20rnew}, that gives an upper limit for tree level $\rho^2$ that is close to the BPS result, should be taken into account, with all the relevant corrections \cite{19rnew}.\par\vskip 1 truecm

\section*{Appendix A. Alternative derivation of the curvature using the BPS condition.}
\hspace*{\parindent} 

In this Appendix we give an alternative derivation of $\tau_{3/2}^{(2)}(1) = 0$, eq. (\ref{54e}), using for $L=2$ excitations the same constraints (translational invariance and equations of motion), supplemented by the BPS condition (\ref{28e}),  that lead to (\ref{44e}) for $L = 1$. \par

This proof is cumbersome, compared to the one of Section 4 using the SR, but illustrates the physical feature that one needs two derivatives in the $L=2$ case. Indeed, with a single derivative one does not obtain any constraint on $\tau_{3/2}^{(2)}$. In general, one would need $L$ derivatives for any $L$, a very involved method compared with the one of Section 5 using the SR.\par

We consider the ${3 \over 2}^-$ doublet, i.e. $L=2$ and $j = {3\over 2}$. The transition matrix element ${1\over 2}^- \to {3 \over 2}^-$ reads~:
$$<D_{3/2}^{(2)}(v') | \overline{h}_{v'}^{(c)} \Gamma h_v^{(b)}|B(v)>\ = \tau_{3/2}^{(2)}(w) Tr \left [ v_{\sigma} \overline{F}_{v'}^{\sigma} \Gamma H_v\right ] \eqno({\rm A.1})$$

We need now {\it two derivatives} to excite the $L=2$ states and obtain a constraint on $\tau_{3/2}^{(2)}(w)$~:
$$i\partial_{\mu} i \partial_{\lambda} <D_{3/2}^{(2)}(v') | \overline{h}_{v'}^{(c)} \Gamma h_v^{(b)}|B(v)>$$
$$= \ <D_{3/2}^{(2)}(v') | \overline{h}_{v'}^{(c)} \left ( \Gamma i \overrightarrow{\cal D}_{\mu}i \overrightarrow{\cal D}_{\lambda} + i \overleftarrow{\cal D}_{\lambda}i \overleftarrow{\cal D}_{\mu}\Gamma +  i \overleftarrow{\cal D}_{\mu}\Gamma i \overrightarrow{\cal D}_{\lambda} + i \overleftarrow{\cal D}_{\lambda}\Gamma i \overrightarrow{\cal D}_{\mu}\right ) h_v^{(b)}|B(v)>$$
$$= \left ( \overline{\Lambda} v_{\mu} - \overline{\Lambda}^* v'_{\mu}  \right )  \left ( \overline{\Lambda} v_{\lambda} - \overline{\Lambda}^* v'_{\lambda}  \right ) <D_{3/2}^{(2)}(1^-)(v') | \overline{h}_{v'}^{(c)} \Gamma h_v^{(b)}|B(v)> \eqno({\rm A.2})$$

\noi where we have kept the same notation $\overline{\Lambda}^*$ as for $L=1$ states, and the different matrix elements are defined by~:
$$ <D_{3/2}^{(2)}(v') | \overline{h}_{v'}^{(c)} \Gamma i \overrightarrow{\cal D}_{\mu}i \overrightarrow{\cal D}_{\lambda} h_v^{(b)}|B(v)>\ = Tr \left [ S_{\sigma\lambda\mu}^{(b,b)} \overline{F}_{v'}^{\sigma}\Gamma H_v \right ]$$
$$ <D_{3/2}^{(2)}(v') | \overline{h}_{v'}^{(c)}  i \overleftarrow{\cal D}_{\lambda}i \overleftarrow{\cal D}_{\mu} \Gamma h_v^{(b)}|B(v)>\ = Tr \left [ S_{\sigma\lambda\mu}^{(c,c)} \overline{F}_{v'}^{\sigma}\Gamma H_v \right ]$$
$$ <D_{3/2}^{(2)}(v') | \overline{h}_{v'}^{(c)}  i \overleftarrow{\cal D}_{\lambda}\Gamma i \overrightarrow{\cal D}_{\mu} h_v^{(b)}|B(v)>\ = Tr \left [ S_{\sigma\lambda\mu}^{(c,b)} \overline{F}_{v'}^{\sigma}\Gamma H_v \right ]$$
$$ <D_{3/2}^{(2)}(v') | \overline{h}_{v'}^{(c)}  i \overleftarrow{\cal D}_{\mu}\Gamma i \overrightarrow{\cal D}_{\lambda} h_v^{(b)}|B(v)>\ = Tr \left [ S_{\sigma\lambda\mu}^{(b,c)} \overline{F}_{v'}^{\sigma}\Gamma H_v \right ]  \eqno({\rm A.3})$$

\noi i.e., one obtains the generalization of (\ref{37e}) to the present case~:
$$S_{\sigma\lambda\mu}^{(b,b)} + S_{\sigma\lambda\mu}^{(b,c)} + S_{\sigma\lambda\mu}^{(c,b)}+ S_{\sigma\lambda\mu}^{(c,c)} = v_{\sigma} \left ( \overline{\Lambda} v_{\mu} - \overline{\Lambda}^* v'_{\mu}  \right )  \left ( \overline{\Lambda} v_{\lambda} - \overline{\Lambda}^* v'_{\lambda}  \right ) \tau_{3/2}^{(2)}\eqno({\rm A.4})$$

\noi The equations of motion give~:
$$\begin{array}{l}Tr \left [ v^{\lambda} S_{\sigma\lambda\mu}^{(b,b)} \overline{F}_{v'}^{\sigma}\Gamma H_v \right ] = 0\\
Tr \left [ v'^{\lambda} S_{\sigma\lambda\mu}^{(c,c)} \overline{F}_{v'}^{\sigma}\Gamma H_v \right ] = 0\\
Tr \left [ v^{\mu} S_{\sigma\lambda\mu}^{(c,b)} \overline{F}_{v'}^{\sigma}\Gamma H_v \right ] = Tr  \left [ v'^{\lambda} S_{\sigma\lambda\mu}^{(c,b)} \overline{F}_{v'}^{\sigma}\Gamma H_v \right ] = 0\\
Tr \left [ v^{\lambda} S_{\sigma\lambda\mu}^{(b,c)} \overline{F}_{v'}^{\sigma}\Gamma H_v \right ] = Tr  \left [ v'^{\mu} S_{\sigma\lambda\mu}^{(b,c)} \overline{F}_{v'}^{\sigma}\Gamma H_v \right ] = 0 \end{array}\eqno({\rm A.5})$$

\noi while the BPS conditions imply~:
$$ \begin{array}{l} <D_{3/2}^{(2)}(1^-)(v') | \overline{h}_{v'}^{(c)} \Gamma i \overrightarrow{\cal D}_{\mu}i \overrightarrow{\cal D}_{\lambda} \gamma^{\lambda} h_v^{(b)}|B(v)>\ = Tr \left [ S_{\sigma\lambda\mu}^{(b,b)} \overline{F}_{v'}^{\sigma}\Gamma\gamma^{\lambda} H_v \right ] = 0\\
 <D_{3/2}^{(2)}(1^-)(v') | \overline{h}_{v'}^{(c)}  i \overleftarrow{\cal D}_{\lambda} \Gamma i\overrightarrow{\cal D}_{\mu} \gamma^{\mu} h_v^{(b)}|B(v)>\ = Tr \left [ S_{\sigma\lambda\mu}^{(c,b)} \overline{F}_{v'}^{\sigma}\Gamma \gamma^{\mu} H_v \right ]=0\\
 <D_{3/2}^{(2)}(1^-)(v') | \overline{h}_{v'}^{(c)}  i \overleftarrow{\cal D}_{\mu}\Gamma i \overrightarrow{\cal D}_{\lambda} \gamma^{\lambda} h_v^{(b)}|B(v)>\ = Tr \left [ S_{\sigma\lambda\mu}^{(b,c)} \overline{F}_{v'}^{\sigma}\Gamma \gamma^{\lambda} H_v \right ] =0\end{array} \eqno({\rm A.6})$$

\noi Notice that there is no condition on $S_{\sigma\lambda\mu}^{(c,c)}$ using the BPS constraint.

The most general parametrisation for $S_{\sigma\lambda\mu}^{(Q,Q')}$ involves 16 terms,
$$\begin{array}{l} S_{\sigma\lambda\mu}^{(Q,Q')} =  \tau_{1}^{(Q,Q')} v_{\sigma}v_{\lambda}v_{\mu} + \tau_{2}^{(Q,Q')} v_{\sigma}v'_{\lambda}v_{\mu} + \tau_{3}^{(Q,Q')} v_{\sigma}v_{\lambda}v'_{\mu} + \tau_{4}^{(Q,Q')} v_{\sigma}v'_{\lambda}v'_{\mu}\\
+ \tau_{5}^{(Q,Q')} v_{\sigma}\gamma_{\lambda}v_{\mu} +\tau_{6}^{(Q,Q')} v_{\sigma} v_{\lambda}\gamma_{\mu} + \tau_{7}^{(Q,Q')} v_{\sigma}v'_{\lambda}\gamma_{\mu} +\tau_{8}^{(Q,Q')} v_{\sigma} \gamma_{\lambda}v'_{\mu}\\
+ \tau_{9}^{(Q,Q')} v_{\sigma}i\sigma_{ \lambda\mu } +\tau_{10}^{(Q,Q')} v_{\sigma} g_{\lambda\mu}\\
+ \tau_{11}^{(Q,Q')} g_{\sigma \lambda}v_{\mu} +\tau_{12}^{(Q,Q')} g_{\sigma \mu}v_{\lambda} +  \tau_{13}^{(Q,Q')} g_{\sigma \lambda}v'_{\mu} +\tau_{14}^{(Q,Q')} g_{\sigma \mu}v'_{\lambda}\\
+ \tau_{15}^{(Q,Q')} g_{\sigma \lambda}\gamma_{\mu} +\tau_{16}^{(Q,Q')} g_{\sigma \mu}\gamma_{\lambda}  \end{array}\eqno({\rm A.7})$$

\noi At $w=1$ one gets, from (A.4)
$$\begin{array}{l} \tau_{1}^{(b,b)} +  \tau_{2}^{(b,b)} +  \tau_{3}^{(b,b)} +  \tau_{4}^{(b,b)}\\
+ \ \tau_{1}^{(b,c)} +  \tau_{2}^{(b,c)} +  \tau_{3}^{(b,c)} +  \tau_{4}^{(b,c)}\\
+ \ \tau_{1}^{(c,b)} + \tau_{2}^{(c,b)} + \tau_{3}^{(c,b)} +  \tau_{4}^{(c,b)}\\
+ \ \tau_{1}^{(c,c)} + \tau_{2}^{(c,c)} +  \tau_{3}^{(c,c)} + \tau_{4}^{(c,c)} = \left ( \overline{\Lambda} - \overline{\Lambda}^*\right )  \tau_{3/2}^{(2)} \end{array} \eqno({\rm A.8})$$

$$ \tau_{i}^{(b,b)} +  \tau_{i}^{(b,c)} +  \tau_{i}^{(c,b)} +  \tau_{i}^{(c,c)} = 0 \qquad (i = 5, \cdots 16) \eqno({\rm A.9})$$

\noi and from the equations of motion (A.5) one gets 
$$\begin{array}{l} \tau_{1}^{(b,b)} +  \tau_{2}^{(b,b)} +  \tau_{10}^{(b,b)} +  \tau_{11}^{(b,b)}-  \tau_{5}^{(b,b)} +  \tau_{9}^{(b,b)} = 0\\
 \tau_{3}^{(b,b)} +  \tau_{4}^{(b,b)} +  \tau_{13}^{(b,b)} -  \tau_{8}^{(b,b)} = 0\\
\tau_{6}^{(b,b)} +  \tau_{7}^{(b,b)} + \tau_{15}^{(b,b)} +  \tau_{9}^{(b,b)} = 0\\
 \tau_{12}^{(b,b)} +  \tau_{14}^{(b,b)}-  \tau_{16}^{(b,b)} = 0 \hbox to 9.5 truecm {} ({\rm A.10}) \\
\\
\tau_{1}^{(c,c)} + \tau_{2}^{(c,c)} +  \tau_{5}^{(c,c)}= 0\\
 \tau_{3}^{(c,c)} + \tau_{4}^{(c,c)} + \tau_{10}^{(c,c)} + \tau_{8}^{(c,c)}  - \tau_{9}^{(c,c)} = 0\\
\tau_{6}^{(c,c)} + \tau_{7}^{(c,c)} + \tau_{9}^{(c,c)} = 0\\
 \tau_{12}^{(c,c)} + \tau_{14}^{(c,c)} + \tau_{16}^{(c,c)} = 0  \hbox to 9.5 truecm {} ({\rm A.11})
 \end{array}$$
 
$$ \begin{array}{l}
\tau_{1}^{(c,b)} + \tau_{3}^{(c,b)} + \tau_{10}^{(c,b)}  + \tau_{12}^{(c,b)} - \tau_{6}^{(c,b)} - \tau_{9}^{(c,b)} =0\\
\tau_{2}^{(c,b)} + \tau_{4}^{(c,b)} + \tau_{14}^{(c,b)} - \tau_{7}^{(c,b)}= 0\\
\tau_{5}^{(c,b)} + \tau_{8}^{(c,b)} + \tau_{16}^{(c,b)} - \tau_{9}^{(c,b)} = 0\\
\tau_{11}^{(c,b)}  + \tau_{13}^{(c,b)} - \tau_{15}^{(c,b)}= 0 \hbox to 9.5 truecm  {} ({\rm A.12})\\
\\
\tau_{1}^{(c,b)} + \tau_{2}^{(c,b)} + \tau_{5}^{(c,b)} = 0\\
 \tau_{3}^{(c,b)} + \tau_{4}^{(c,b)} + \tau_{10}^{(c,b)} + \tau_{8}^{(c,b)} - \tau_{9}^{(c,b)}=0\\
\tau_{6}^{(c,b)} + \tau_{7}^{(c,b)} + \tau_{9}^{(c,b)} =0\\
 \tau_{12}^{(c,b)} + \tau_{14}^{(c,b)} + \tau_{16}^{(c,b)} = 0  \hbox to 9.5 truecm {} ({\rm A.13})\\
\\
\tau_{1}^{(b,c)} +  \tau_{2}^{(b,c)} +  \tau_{10}^{(b,c)} +  \tau_{11}^{(b,c)} - \tau_{5}^{(b,c)} +  \tau_{9}^{(b,c)} = 0\\
\tau_{3}^{(b,c)} +  \tau_{4}^{(b,c)}+ \tau_{13}^{(b,c)} - \tau_{8}^{(b,c)} = 0\\
\tau_{6}^{(b,c)} +  \tau_{7}^{(b,c)}+ \tau_{15}^{(b,c)} +  \tau_{9}^{(b,c)}= 0\\
\tau_{12}^{(b,c)} +  \tau_{14}^{(b,c)} - \tau_{16}^{(b,c)} = 0   \hbox to 9.5 truecm {}({\rm A.14})\\
\\
\tau_{1}^{(b,c)} +  \tau_{3}^{(b,c)} +  \tau_{6}^{(b,c)} =0\\
\tau_{2}^{(b,c)} + \tau_{4}^{(b,c)} +  \tau_{10}^{(b,c)} + \tau_{7}^{(b,c)} +  \tau_{9}^{(b,c)}= 0\\
\tau_{5}^{(b,c)} + \tau_{8}^{(b,c)} - \tau_{9}^{(b,c)} = 0\\
\tau_{11}^{(b,c)} +  \tau_{13}^{(b,c)} - \tau_{15}^{(b,c)} = 0 \hbox to 9.5 truecm {} ({\rm A.15})
\end{array}$$

\noi while the BPS conditions (A.6) imply
$$\begin{array}{l} \tau_{1}^{(b,b)} +  \tau_{2}^{(b,b)} -  4\tau_{5}^{(b,b)} +  2\tau_{9}^{(b,b)}+  2\tau_{10}^{(b,b)} +  2\tau_{11}^{(b,b)} = 0\\
 \tau_{3}^{(b,b)} +  \tau_{4}^{(b,b)} - 2  \tau_{7}^{(b,b)} -  4\tau_{8}^{(b,b)} + 2 \tau_{13}^{(b,b)}= 0\\
\tau_{6}^{(b,b)} +  3\tau_{7}^{(b,b)} + 3\tau_{9}^{(b,b)} -  \tau_{10}^{(b,b)} + 2 \tau_{15}^{(b,b)}= 0\\
 \tau_{12}^{(b,b)} +  \tau_{14}^{(b,b)}- 2 \tau_{15}^{(b,b)} - 4 \tau_{16}^{(b,b)}= 0\\
 \tau_{1}^{(b,b)} +  \tau_{2}^{(b,b)} +  2\tau_{5}^{(b,b)} = 0\\
 \tau_{3}^{(b,b)}+  \tau_{4}^{(b,b)} +  2\tau_{7}^{(b,b)} + 2\tau_{8}^{(b,b)} = 0\\
 \tau_{6}^{(b,b)} -  \tau_{7}^{(b,b)} -  \tau_{9}^{(b,b)} + \tau_{10}^{(b,b)} = 0\\
\tau_{12}^{(b,b)} + \tau_{14}^{(b,b)} + 2 \tau_{15}^{(b,b)} + 2 \tau_{16}^{(b,b)} = 0 \end{array} \eqno({\rm A.16})$$ 

$$\begin{array}{l}\tau_{1}^{(c,b)} + \tau_{3}^{(c,b)} -4 \tau_{6}^{(c,b)}  -2 \tau_{9}^{(c,b)} +2 \tau_{10}^{(c,b)} +2 \tau_{12}^{(c,b)} =0\\
\tau_{2}^{(c,b)} + \tau_{4}^{(c,b)} - 2 \tau_{8}^{(c,b)} - 4\tau_{7}^{(c,b)}+ 2\tau_{14}^{(c,b)} =0\\
 \tau_{5}^{(c,b)} -3 \tau_{9}^{(c,b)} - \tau_{10}^{(c,b)} + 3\tau_{8}^{(c,b)}  + 2\tau_{16}^{(c,b)} =0\\
\tau_{11}^{(c,b)}+\tau_{13}^{(c,b)} -4 \tau_{15}^{(c,b)} -2 \tau_{16}^{(c,b)} = 0\\
 \tau_{1}^{(c,b)} + \tau_{3}^{(c,b)} + 2\tau_{6}^{(c,b)} = 0\\
 \tau_{2}^{(c,b)} + \tau_{4}^{(c,b)} + 2\tau_{7}^{(c,b)} +2 \tau_{8}^{(c,b)} =0\\
\tau_{5}^{(c,b)} - \tau_{8}^{(c,b)} + \tau_{9}^{(c,b)} + \tau_{10}^{(c,b)} = 0\\
\tau_{11}^{(c,b)} + \tau_{13}^{(c,b)} + 2\tau_{15}^{(c,b)} +2 \tau_{16}^{(c,b)} = 0 \end{array}\eqno({\rm A.17})$$
\vskip 5 truemm

$$\begin{array}{l}\tau_{1}^{(b,c)} +  \tau_{2}^{(b,c)} -4  \tau_{5}^{(b,c)} +  2\tau_{9}^{(b,c)} +2 \tau_{10}^{(b,c)} +  2\tau_{11}^{(b,c)} = 0\\
\tau_{3}^{(b,c)} +  \tau_{4}^{(b,c)}-2 \tau_{7}^{(b,c)} - 4\tau_{8}^{(b,c)} + 2\tau_{13}^{(b,c)} = 0\\
 \tau_{6}^{(b,c)}+ 3\tau_{7}^{(b,c)} +  3\tau_{9}^{(b,c)} - \tau_{10}^{(b,c)} + 2 \tau_{15}^{(b,c)} = 0\\
\tau_{12}^{(b,c)} + \tau_{14}^{(b,c)} -2  \tau_{15}^{(b,c)} -4  \tau_{16}^{(b,c)} =0\\
\tau_{1}^{(b,c)} + \tau_{2}^{(b,c)} + 2 \tau_{5}^{(b,c)} = 0\\
\tau_{3}^{(b,c)} +  \tau_{4}^{(b,c)} + 2\tau_{7}^{(b,c)} + 2\tau_{8}^{(b,c)} = 0\\
\tau_{6}^{(b,c)} - \tau_{7}^{(b,c)} -  \tau_{9}^{(b,c)} + \tau_{10}^{(b,c)} = 0\\
\tau_{12}^{(b,c)} + \tau_{14}^{(b,c)} +2  \tau_{15}^{(b,c)} + 2\tau_{16}^{(b,c)} =0 \end{array}\eqno({\rm A.18})$$

\noi There are other equations following from the symmetries of (A.3) and the definitions (A.7)~:
$$\begin{array}{llll}
\tau_{1}^{(b,c)} = \tau_{1}^{(c,b)} &\quad \tau_{4}^{(b,c)} = \tau_{4}^{(c,b)} &\quad\tau_{9}^{(b,c)} = - \tau_{9}^{(c,b)} &\quad\tau_{10}^{(b,c)} = \tau_{10}^{(c,b)}  \\
\tau_{2}^{(b,c)} = \tau_{3}^{(c,b)} &\quad\tau_{3}^{(b,c)} = \tau_{2}^{(c,b)} &\quad\tau_{5}^{(b,c)} = \tau_{6}^{(c,b)} &\quad\tau_{6}^{(b,c)} = \tau_{5}^{(c,b)}  \\
\tau_{7}^{(b,c)} = \tau_{8}^{(c,b)} &\quad\tau_{8}^{(b,c)} = \tau_{7}^{(c,b)} &\quad\tau_{11}^{(b,c)} =  \tau_{12}^{(c,b)} &\quad\tau_{12}^{(b,c)} = \tau_{11}^{(c,b)}  \\
\tau_{13}^{(b,c)} = \tau_{14}^{(c,b)} &\quad\tau_{14}^{(b,c)} = \tau_{13}^{(c,b)} &\quad\tau_{15}^{(b,c)} =  \tau_{16}^{(c,b)} &\quad\tau_{16}^{(b,c)} = \tau_{15}^{(c,b)} 
 \hskip 2 truecm ({\rm A.19})\\
 \\
\tau_{2}^{(b,b)} = \tau_{3}^{(b,b)} &\quad\tau_{5}^{(b,b)} = \tau_{6}^{(b,b)} &\quad\tau_{7}^{(b,b)} = \tau_{8}^{(b,b)} &  \\
\tau_{11}^{(b,b)} = \tau_{12}^{(b,b)} &\quad\tau_{13}^{(b,b)} = \tau_{14}^{(b,b)} &\quad\tau_{15}^{(b,b)} =  \tau_{16}^{(b,b)}&
 \hskip 4.7 truecm({\rm A.20})\\
\\
\tau_{2}^{(c,c)} = \tau_{3}^{(c,c)} &\quad\tau_{5}^{(c,c)} = \tau_{6}^{(c,c)} &\quad\tau_{7}^{(c,c)} = \tau_{8}^{(c,c)}   &\\
\tau_{11}^{(c,c)} = \tau_{12}^{(c,c)} &\quad\tau_{13}^{(c,c)} = \tau_{14}^{(c,c)} &\quad\tau_{15}^{(c,c)} = \tau_{16}^{(c,c)}&
 \hskip 4.7 truecm({\rm A.21}) \end{array}$$

A careful study of the system of linear equations (A.8)-(A.21), among other relations, implies
$$\tau_{3/2}^{(2)}(1) = 0 \eqno({\rm A.22})$$

\noi as we wanted to demonstrate, and therefore (\ref{46e}) follows.

\section*{Acknowledgement}
\hspace*{\parindent} This work was supported by the EC contract HPRN-CT-2002-00311 (EURIDICE).

\end{document}